\documentclass[12pt]{article}
\usepackage{amsmath, amsthm, amssymb}
\usepackage{epsfig}
\begin{document}
\baselineskip= 24pt
\begin{center}
{\Large  Thermal Degradation of Adsorbed Bottle-Brush Macromolecules:\\ A
Molecular Dynamics Simulation }\\
\vskip 0.2 true cm
\footnote{To whom correspondence should be addressed. E-mail:
milchev@mail.uni-mainz.de}{Andrey Milchev$^{1,2}$, Jaroslaw Paturej$^{1,3}$,
Vakhtang G. Rostiashvili$^1$, and Thomas A. Vilgis$^1$} \\
\vskip 0.2true cm
{$^1$ Max Planck Institute for Polymer Research
10 Ackermannweg, 55128 Mainz, Germany\\
$^2$ Institute for Physical Chemistry, Bulgarian Academy of Science,
1113 Sofia, Bulgaria\\
$^3$ Institute of Physics, University of Szczecin, Wielkopolska 15,
70451 Szczecin, Poland}
\end{center}
\date{\today}

\begin{abstract}
The scission kinetics of bottle-brush molecules in solution and on an adhesive
substrate is modeled by means of Molecular Dynamics simulation with Langevin
thermostat. Our macromolecules comprise a long flexible polymer backbone with
$L$ segments, consisting of breakable bonds, along with two side chains of
length $N$, tethered to each segment of the backbone. In agreement with recent
experiments and theoretical predictions, we find that bond cleavage is
significantly enhanced on a strongly attractive substrate even though the
chemical nature of the bonds remains thereby unchanged.

We find that the mean bond life time $\langle \tau \rangle$ decreases upon
adsorption by more than an order of magnitude even for brush molecules with
comparatively short side chains $N=1 \div 4$. The distribution of scission
probability
along the bonds of the backbone is found to be rather sensitive regarding the
interplay between length and grafting density of side chains. The life time
$\langle \tau \rangle$ declines with growing contour length $L$ as  $\langle
\tau \rangle \propto L^{-0.17}$, and with side chain length as $\langle \tau
\rangle \propto N^{-0.53}$. The probability distribution of fragment lengths at
different times agrees well with experimental observations.  The variation of
the mean length $L(t)$ of the fragments with elapsed time confirms the notion of
the thermal degradation process as a first order reaction.

\end{abstract}

%\pacs{82.35.Lr, 82.35.Jk, 87.15.Aa}

%\maketitle

\section{Introduction}

The study of degradation and stabilization of polymers is important both from
practical and theoretical viewpoints \cite{Allen}. Disposal of plastic wastes
has grown rapidly to a world problem so that increasing environmental concerns
have prompted researchers to investigate plastics recycling by degradation as an
alternative \cite{Madras}. On the other hand,  degradation of polymers in
different environment is a major limiting factor in their application. Recently,
with the advent of exploiting biopolymers as functional materials
\cite{Schulten,Han}, the stability of  such materials has become an issue of
primary concern.

Most theoretical investigations of polymer degradation have focused  so far on
determining the rate of change of average molecular weight
\cite{Simha,Jellinek,Ballauff,Ziff,Cheng,Nyden2,Doerr,Wang,Hathorn,Doruker}. The
main assumptions of the theory are that each link in a long chain molecule has
equal strength and equal accessibility, that they are broken at random, and that
the probability of rupture is proportional to the number of links present.
Experimental
studies of polystyrene, however, have revealed discrepancies \cite{Jellinek}
with the theory \cite{Simha} so, for example, the thermal degradation stops
completely or slows down markedly when a certain chain length is reached. Only
few theoretic studies \cite{Lee,Sokolov} have recently explored how does the
single polymer chain dynamics  affect the resulting bond rupture probability. In
both studies, however, for the sake of theoretical tractability one has worked
with a model of a Gaussian chain bonded by linear (harmonic) forces whereby the
anharmonic (non-linear) nature of the bonding interactions was not taken into
account. One could claim that the process of thermal degradation still remains
insufficiently studied and understood.

Meanwhile, recently it was found experimentally \cite{Maty,Lebed,Park} that
covalent bonds may spontaneously break upon adsorption of brush-like
macromolecules onto a substrate. One studied brushes consisting of a
poly(2-hydroxyethyl metacrylate) backbone and a poly($N$-butyl acrylate) (PAB)
side chains with degrees of polymerization $L =  2150 \pm 100$ and $N = 140 \pm
5$, and found spontaneous rupture of covalent bonds (which are otherwise hard to
break) upon adsorption of these molecules on mica, graphite, or water-propanol
interfaces \cite{Maty}. As the densely grafted side chains adsorb, they
experience steric repulsion due to monomer crowding which creates tension in the
backbone. This tension, which depends on the grafting density, the side chain
length, and the extent of substrate attraction, effectively lowers the energy
barrier for dissociation, decreasing the bond life time \cite{Beyer}.  Thus, one
may observe amplification of bond tension from the pico-newton to nano-newton
range which facilitates thermal degradation considerably.

Also recently, in several works Panyukov and collaborators \cite{Panyuk,Pan_PRL}
predicted and theoretically described the effect of tension amplification in
branched macromolecules. They argued that the brush-like architecture allows
focusing of the side chain tension to the backbone whereby at given temperature
$T$ the tension in the backbone becomes proportional to the length of the side
chain, $f \approx f_0 N$ \cite{Panyuk,Pan_PRL}. The maximum tension in the side
chains is $f_0 \approx k_BT / b$ with $k_B$ -being the Boltzmann constant, and
$b$ - the Kuhn length (or, the monomer diameter for absolutely flexible chains).

The effect of adsorption-induced bond scission might have important implication
for surface chemistry, in general,  and for specific applications of new macro-
and supramolecular materials, in particular,  for example, by steering the
course of chemical reactions. One may use adsorption as a convenient way to
exceed the strength of covalent bonds and invoke irreversible fracture of
macromolecules, holding the key to making molecular (DNA) architectures that
undergo well-defined fragmentation upon adsorption.

In the present  investigation we explore the process of chain fragmentation in
desorbed and adsorbed bottle-brush macromolecules by means of a coarse-grained
bead-spring model and Langevin dynamics. In Section \ref{Model} we describe
briefly our model and then present our simulation results in Section
\ref{MD_results}. A summary of our results and conclusions is presented in
Section \ref{Conclusions}. Anticipating, one might claim that the reported
results appear in good agreement with observations and theoretical predictions.

\section{The Model} \label{Model}

We consider a $3d$ coarse-grained model of a polymer chain which consists of $L$
repeatable units (monomers) connected by bonds, whereby each bond of length
$b$ is described by a Morse potential,
\begin{equation}\label{Morse_pot}
U_M(r) = D \{1 - \exp[- \alpha (r - b) \}^2
\end{equation}
with a parameter $\alpha \equiv 1$. The dissociation energy of such bonds is
$D$, measured in units of $k_BT$, where $k_B$ denotes the Boltzmann constant and
$T$ is the temperature. The maximum restoring force of the Morse potential,
$f_{max} = -dU_M/dr = \alpha D / 2$, is reached at the inflection point, $r =
b+\alpha^{-1} \ln(2)$. This force $f_{max}$  determines the tensile strength of
the chain. Since the bond extension $r-b$ between nearest-neighbor
monomers along the polymer backbone  in our $3d$-model is always positive, the
Morse potential Eq.~(\ref{Morse_pot})  is only weakly repulsive and segments
could partially penetrate one another at $r < b$. Therefore, in order to allow
properly for the {\em excluded volume} interactions between bonded monomers, we
take the bond potential as a sum of $U_M(r)$ {\em and}  the so called
Weeks-Chandler-Anderson (WCA) potential, $U_{WCA}(r)$, (i.e., the shifted and
truncated repulsive branch of the Lennard-Jones potential);
\begin{equation}\label{WCA_pot}
U_{WCA}(r) = 4\epsilon \left [\left( \frac{\sigma}{r}\right )^{12} -
\left( \frac{\sigma}{r}\right )^{6} + \frac{1}{4}\right ] \Theta(2^{1/6}
\sigma - r))
\end{equation}
with $\Theta(x) = 0\; \mbox{or}\; 1$ for $x < 0$ or $x \ge 0$, and $\epsilon =
1$. The non-bonded interactions between monomers are also taken into account by
means of the WCA potential, Eq.~(\ref{WCA_pot}). Thus the interactions in our
model correspond to good solvent conditions. The length scale is set by the
parameter $\sigma = 1$ whereby the monomer diameter $b = 2^{1/6} \sigma \approx
1.12 \sigma$.

In our MD simulation we use a Langevin equation, which describes the Brownian
motion of a set of interacting particles whereby the action of the solvent is
split into slowly evolving viscous force and a rapidly fluctuating  stochastic
force:
\begin{equation}\label{Langevin_eq}
 m \overrightarrow{\dot{v}}_i(t) = - \zeta \vec{v}_i + \vec{F}_M^i(t) +
\vec{F}_{WCA}^i(t) + \vec{R^i}(t) .
\end{equation}
The random force which represents the incessant collisions of the monomers with
the solvent molecules satisfy the fluctuation-dissipation theorem $\langle
R^i_{\alpha}(t) R^j_{\beta}(t') \rangle = 2\zeta k_B T \delta_{ij}
\delta_{\alpha \beta}\delta(t-t')$. The friction coefficient $\zeta$ of the
Langevin thermostat, used for equilibration, has been set at $0.25$. The
integration step is $0.002$ time units (t.u.) and time is measured in units of
$\sqrt{m/\sigma^2 D}$ where $m$ denotes the mass of the beads, $m=1$. We
emphasize at this point that in our coarse-grained modeling no explicit solvent
particles are included.
\begin{figure}[ht]
\begin{center}
\includegraphics[scale=0.42]{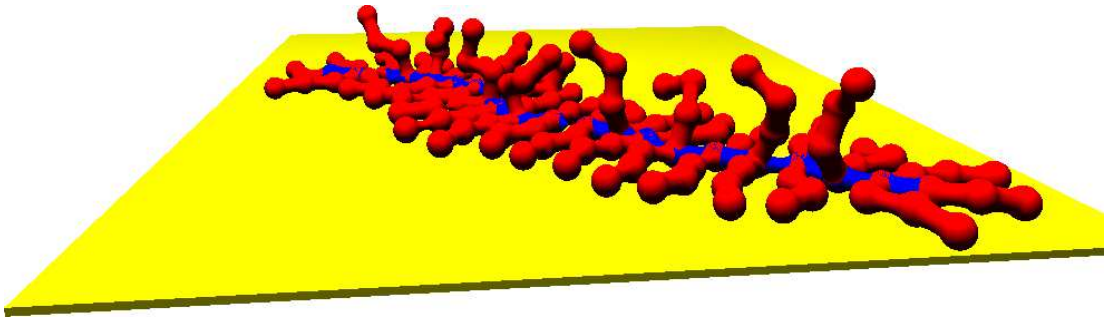}
\caption{A snapshot of a bottle-brush molecule (a ``centipede``) with $L=20$
backbone monomers (blue) and $122$ side chains (red) of length $N=4$. Here $k_BT
= 1$ and strength of adsorption $\epsilon_s = 9.5$. Side chains which are too
strongly squeezed by the neighbors are seen occasionally to get off the
substrate in order to minimize free energy.}
\label{snap_N20S3}
\end{center}
\end{figure}

Two side chains of length $N$ are tethered to {\em each} repeatable unit of the
backbone except for the terminal beads of the polymer backbone where a single
side chains is anchored only. Thus the total number of monomers in the
bottle-brush macromolecule is $M = L + 2N(L+1)$. Because of the high grafting
density, we use rather short side chains $N = 1 \div 5$ in our simulations -
Fig.~\ref{snap_N20S3}.

For the bonded interaction in the side chains we take the frequently used
Kremer-Grest potential, $U_{KG}(r) = U_{WCA}(r) + U_{FENE}(r)$, with the
so-called 'finitely-extensible non-linear elastic' (FENE) potential,
\begin{equation} \label{FENE}
 U_{FENE}(r) = - \frac{1}{2} k r_0^2 \ln \left[ 1 - \left ( \frac{r}{r_0}
 \right ) ^2 \right].
\end{equation}
In Eq.~(\ref{FENE})  $k = 30, \; r_0 = 1.5$, so that the total potential
$U_{KG}(r)$ has a minimum at bond length $r_{bond} \approx 0.96$. Thus, the
bonded interaction, $U_{KG}(r)$, makes the bonds of the side chains in our model
unbreakable whereas those of the backbone may and do undergo scission.

The substrate in the present study is considered simply as a structureless
adsorbing plane, with a Lennard-Jones potential acting with strength
$\epsilon_s$ in the perpendicular $z-$direction.  In our simulations we consider
as a rule the case of {\em strong} adsorption, $\epsilon_s / k_BT = 5.0 \div
10.0$.

We start the simulation with a well equilibrated conformation of the chain and
examine the thermal scission of the bonds. We measure the mean life time $\tau$
until the first bond rupture occurs, and average these times over more than
$2\times 10^4$ events so as to determine the mean $\langle \tau \rangle$ which
is also referred to as Mean First Breakage Time (MFBT). In the course of the
simulation we also sample the probability distribution of bond breaking
regarding their position in the chain (a rupture probability histogram), the
probability distribution of the First Breakage Time, $\tau$, as well as other
quantities of interest. At periodic intervals we analyze the length distribution
of backbone fragments and establish the Probability Distribution Function (PDF)
of fragment sizes, $P(n,t)$, which also yields the time evolution of the mean
fragment length $L(t)$.

Since in the problem of thermal degradation there is no external force acting
on the chain ends, a well defined activation barrier for a bond scission is
actually missing, in contrast to the case of applied tensile force. Therefore, a
definition of an unambiguous criterion for  bond breakage is not self-evident.
Moreover, depending on the degree of stretching, bonds may break and then
recombine again. Therefore, in our numeric experiments we use a sufficiently
large expansion of the bond, $r_h = 2 b$, as a threshold to a broken state of
the bond. This convention is based on our checks that the probability for
recombination of bonds, stretched beyond $r_h$, is sufficiently small.

\section{Simulation Results}
\label{MD_results}

\subsection{Equilibrium Properties} \label{structure}

\begin{figure}[ht]
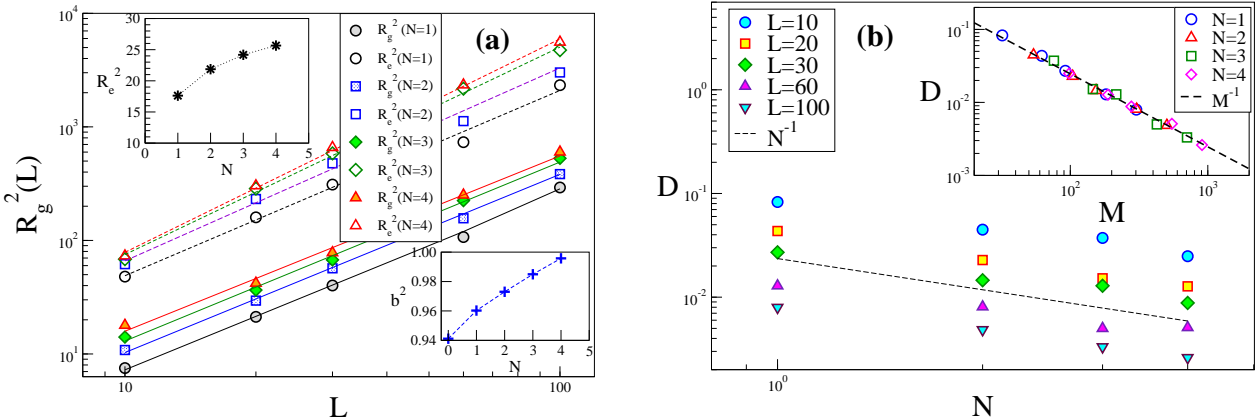

\vspace{0.5cm}
\begin{center}
\hspace{-1.5cm}
\includegraphics[scale=0.32]{Rg_N.eps}
\hspace{0.5cm}
\includegraphics[scale=0.44]{D.eps}
\caption{(a)  Variation of the gyration radius $R_g^2$ and the
end-to-end distance $R_e^2$ with the degree of polymerization $L$ in a brush
molecule for side chains of length $N$. Lines denote a scaling relationship
$R_g^2 \propto R_e^2 \propto L^{2\nu}$. Insets show the increase of $R_e^2$  and
the mean squared bond length $b^2$ with changing side chain length $N$ for
$L=30$.  (b) Diffusion coefficient $D$ vs $N$ for brush molecules of
different length $L$. In the inset $D$ is plotted against the total number of
monomers in the bottle-brush macromolecule $M = L + 2N(L+1)$. The dashed
straight line indicates a $D \propto N^{-1}$ power law.}
\label{struc}
\end{center}
\end{figure}

We have checked typical properties of the strongly adsorbed brush molecules as
the variation of the mean radius of gyration $R_g^2$ and the mean end-to-end
distance between terminal points on the  polymer backbone, $R_e^2$, for several
lengths of the side chains, $N$ - see Fig.~\ref{struc}a. One can easily verify
from  Fig.~\ref{struc}a, that the structure of the bottle-brush indicates a
typical quasi-$2d$ behavior, as one would expect for the case of strong
adsorption.
One observes a scaling behavior $R_g^2 \propto L^{2\nu}$ where the power-law
Flory exponent attains a value $\nu = 0.76 \pm 0.01$ that is  close to the exact
one, $\nu_{2d} = 3/4$. From the insets in Fig.~\ref{struc} one can see that the
end-to-end distance , $R_e^2$, of the backbone steadily increases with growing
length $N$ of the side chains. The same applies for the mean bond length $b^2$
between segments along the backbone as function of $N$.  Evidently, due to the
high grafting density the side chains repel and stretch each other into an
extended conformation. The steric repulsion between side chains is strongly
enhanced when the macromolecule is adsorbed and attains a quasi-twodimensional
conformation. As a result, both its contour and persistent lengths are
increased.

\subsection{Scission Probability Histogram}

\begin{figure}[ht]
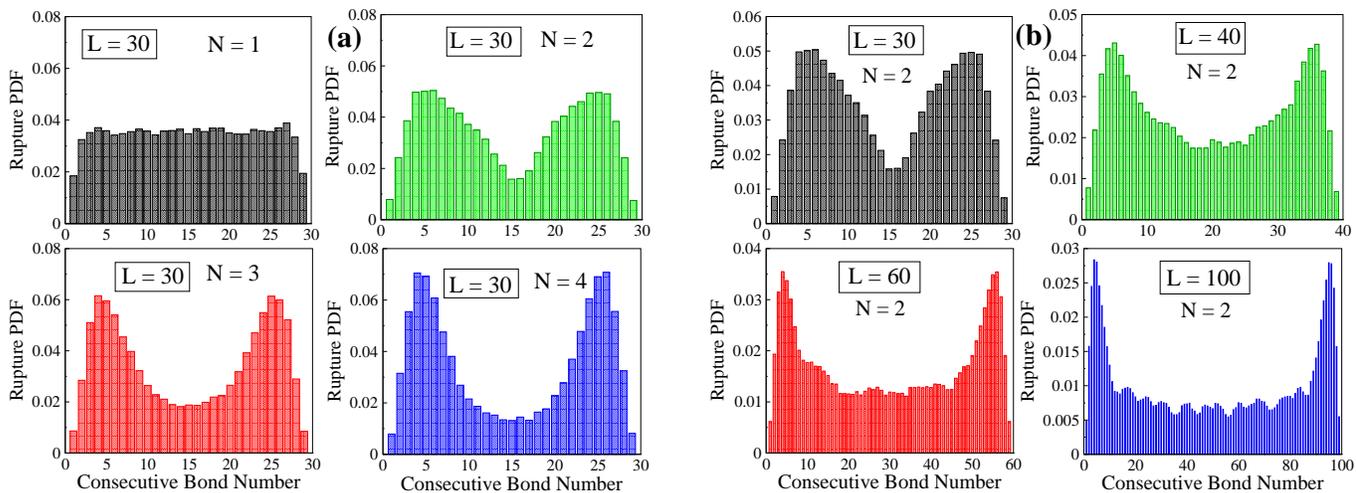

\vspace{0.5cm}
\begin{center}
\hspace{-1.5cm}
\includegraphics[scale=0.32]{rhN30_L.eps}
\hspace{0.5cm}
\includegraphics[scale=0.32]{rhN_L2.eps}
\caption{(a) Scission probability histogram for a polymer backbone with $L=30$
and different length of the side chains $N$.  (b) Variation of the scission
probability histogram with contour length $L$ for brush molecules with fixed
side chain length $N=2$.}
\label{histo}
\end{center}
\end{figure}

We examine the distribution of scission probability (the probability of bond
rupture)
along the polymer backbone for the case of a strong adsorption, $T = 0.10,\;
\epsilon_s = 0.5$ in Fig.~\ref{histo}. One can readily verify from
Fig.~\ref{histo}a that for a given contour length $L$ the shape of the
probability
histogram changes qualitatively as the length of side chains $N$ is increased
beyond one, $N > 1$.
While for $N=1$ the scission probability is uniformly distributed along the
backbone (being significantly diminished only in the vicinity of both terminal
bonds), for $N > 1$, in contrast, one observes a well expressed minimum in the
probability in the middle of the chain in between the two pronounced maxima
(''horns'')
close to the chain ends. This effect persists and is even enhanced as the
contour length $L$ gets larger - Fig.~\ref{histo}b. Occasionally, some
additional maxima show up in longer molecules, $L > 30$, which are then found to
disappear with improved statistics. We have tracked down these temporal maxima
in the scission probabilities as corresponding to local bends and kinks in the
contour
which, owing to mutual squeezing or rarefaction, strongly change the side chains
mobility and, presumably, the induced tension. Since conformations of adsorbed
long molecules change rather slowly, a very large number of simulation runs is
needed before such spurious maxima disappear.

Evidently, with growing length $N$ of the side chains the minimum gets deeper
and broader, indicating that breakage happens most frequently in bonds which are
very close to the terminal bonds of the backbone. Given the high density of
grafting,  one may interpret this effect as a consequence  of the mutual
immobilization and blocking of the side chains which are placed in the middle.
In comparison, side chains that are closer to the ends of the polymer backbone
still have sufficient freedom to move and, therefore, contribute locally to
stronger tension in the vicinity of both ends of the backbone. Side chains of
length $N=1$, on the other hand, are too short to block one another. Therefore
brush molecules with $N=1$ behave similar to such with longer side chains but
at sufficiently lower grafting density.

\subsection{Dependence of $\langle \tau \rangle$ on $L$}

\begin{figure}[ht]
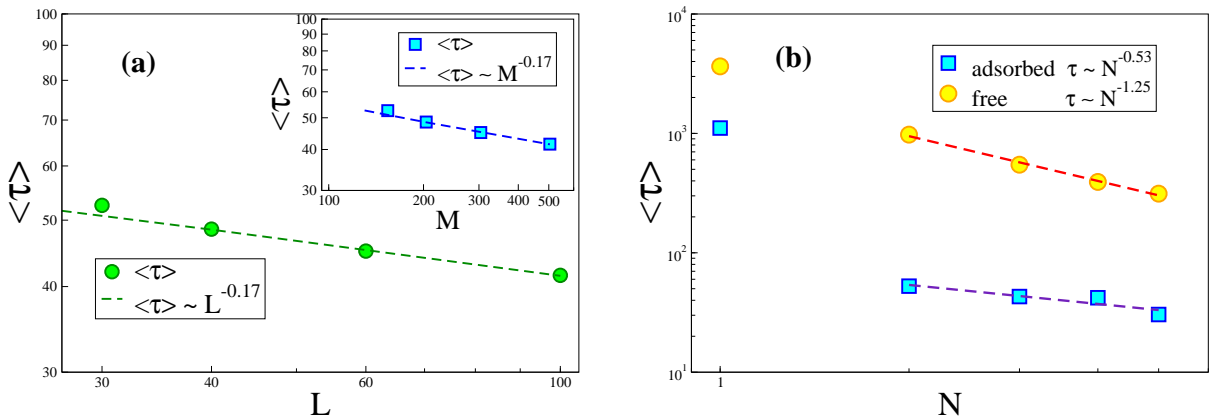

\vspace{0.5cm}
\begin{center}
\hspace{-1.5cm}
\includegraphics[scale=0.32]{tauN_L2.eps}
\hspace{0.5cm}
\includegraphics[scale=0.32]{tauN30_L.eps}
\caption{(a) Variation of the MFBT $\langle \tau \rangle$ with contour length
$L$ and with total number of monomers $M$ of the brush molecule (inset) for
length of the  side chains $N = 2$. Here $k_BT = 0.10$. (b) Mean life time
$\langle \tau \rangle$ vs $N$ for a desorbed (free) and adsorbed brush molecule
with $L=30$.}
\label{tau_L}
\end{center}
\end{figure}

In Fig.~\ref{tau_L}a we show the dependence  of the mean time before any of the
backbone bonds breaks, i.e., the MFBT $\langle \tau \rangle$ on the contour
length $L$ and on the total number of segments in the bottle-brush molecule $M =
L+ 2 N (L+1)$. Evidently, one observes a well expressed power law, $\langle \tau
\rangle \propto L^{-\alpha}$ with exponent $\alpha \approx 0.17$. Since for
large $L$ one has $M \propto L$, the variation of  $\langle \tau \rangle$ with
$M$ is the same.
This finding is important because it indicates that $\langle \tau \rangle$
depends rather weakly on the total number of bonds that might break, in clear
contrast to thermal degradation of polymers without side chains \cite{Jarek}
where $\alpha = 1$. Indeed, when bonds break uncorrelated and entirely at
random, the probability that {\em any} of the $L$ bonds may undergo scission
within a certain time interval should be  proportional to the total number of
bonds, and therefore $\langle \tau \rangle \propto 1 / L$. In cases of chain
scission when  a constant external force pulls at the ends of the polymer,
however, one finds typically $\alpha < 1$ \cite{Ghosh} whereby  the value of
$\alpha$ steadily decreases as the force strength grows. This suggests a gradual
crossover from a predominantly individual to a more concerted mechanism of bond
scission. In adsorbed bottle-brush molecules it is the side chains that induce
tension in the polymer backbone and thus lead to rupture behavior similar to
that with external force.

In Fig.~\ref{tau_L}b we compare the dependence of $\langle \tau \rangle$ on
length $N$ of the side chains, comparing non-adsorbed (free) and adsorbed brush
molecules of length $L = 30$. As far as the side chains are rather short, one
should not overestimate the observed power-law dependence, with $\langle \tau
\rangle \propto N^{-1.25}$ for free, and $\langle \tau \rangle \propto
N^{-0.53}$ for strongly adsorbed molecules. Generally, adsorption is found to
diminish the mean rupture time by more than an order of magnitude alone, at
least for $N>1$. As mentioned before, the case $N=1$ where neighboring side
chains almost do not overlap is qualitatively different so, upon adsorption,
the MFBT shortens by a factor of three  only.

\subsection{Fragment Size Distribution}

In the present work we studied the fragmentation kinetics and the resulting
molecular weight distribution, $P(n,t)$, of  strongly adsorbed
\begin{figure}[ht]
\vspace{0.5cm}
\begin{center}
\hspace{-1.5cm}
\includegraphics[scale=0.32]{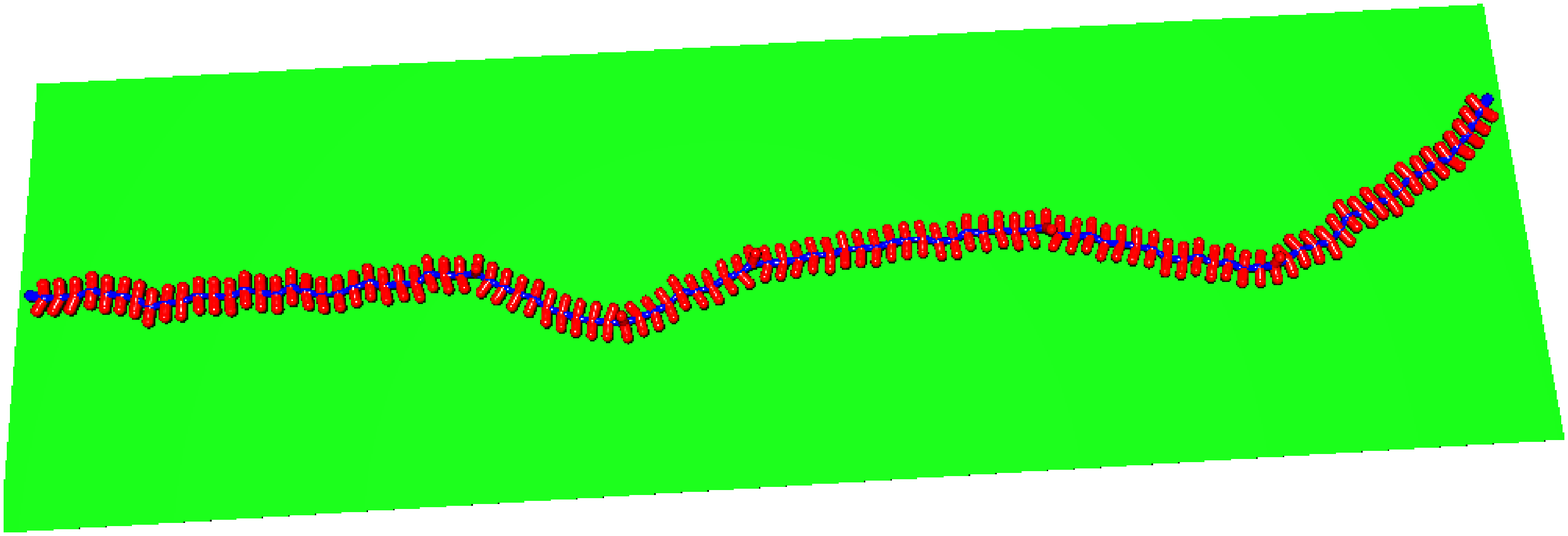}
\hspace{0.5cm}
\includegraphics[scale=0.32]{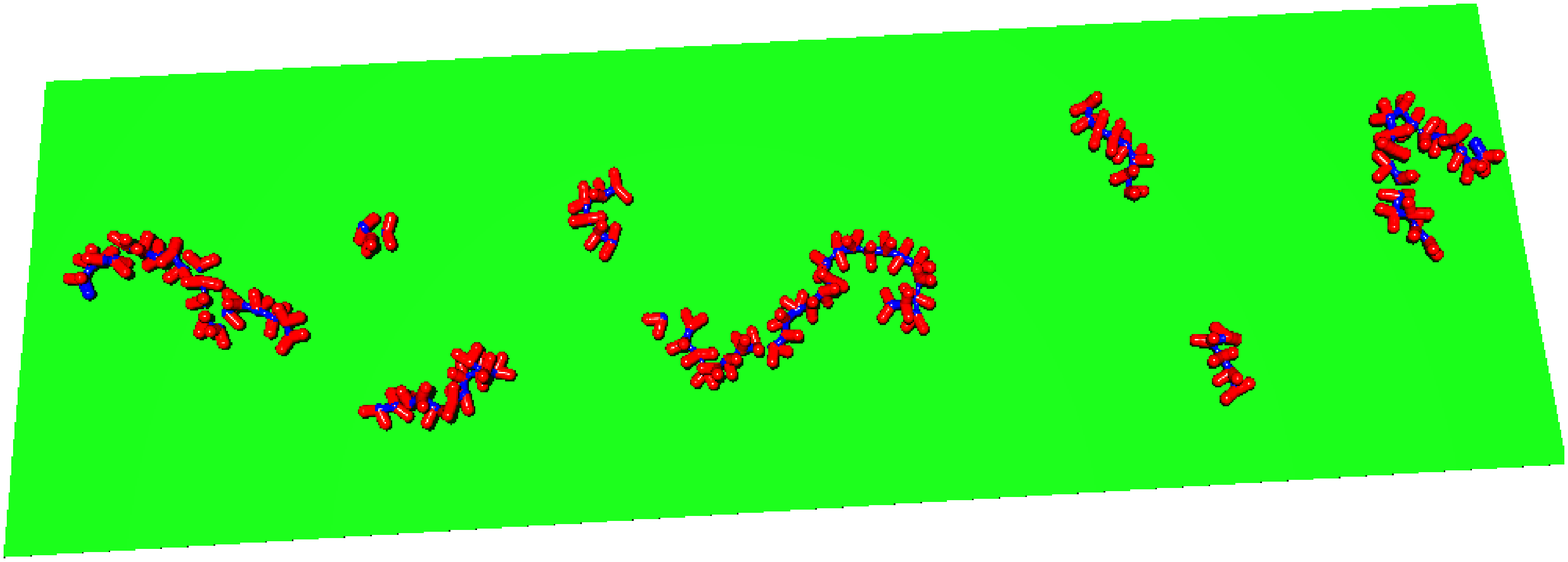}
\caption{Snapshots of an adsorbed bottle-brush macromolecule (a ``centipede'')
with backbone length $L=100$ and side length $N=1$ at $T = 0.12$ and $\epsilon_s
= 0.50$ before (above) and after (below) the fragmentation process is completed
at $t = 600\; t. u.$}
\label{snap_broken}
\end{center}
\end{figure}
bottle-brush molecules for the shortest side chains $N=1$, - see
Fig.~\ref{snap_broken} - since, as shown above, they do not mutually overlap
very strongly and produce a scission probability distribution for the bonds
along the polymer backbone that matches the one, inferred from experiment
\cite{Maty}. If one assumes that the scission kinetics is described by a
first-order reaction, then one may describe the decrease in the average contour
length of the fragments with elapsed time \cite{Lebed,Park} as
\begin{equation} \label{L_ave}
 \left ( \frac{1}{L(t)} - \frac{1}{L_{\infty}} \right ) =  \left (
\frac{1}{L_0} - \frac{1}{L_{\infty}} \right ) e^{-k t},
\end{equation}
where $L_0$ is the initial contour length at $t=0$ and $L_{\infty}$ is the mean
contour length of polymer chains at infinite time. The lower limit of the chain
length of fractured molecules is ascribed to reduction of the backbone tension
as molecular brushes with short backbones adopt a more relaxed conformation.
\begin{figure}[ht]
\vspace{0.5cm}
\begin{center}
\hspace{-1.5cm}
\includegraphics[scale=0.7]{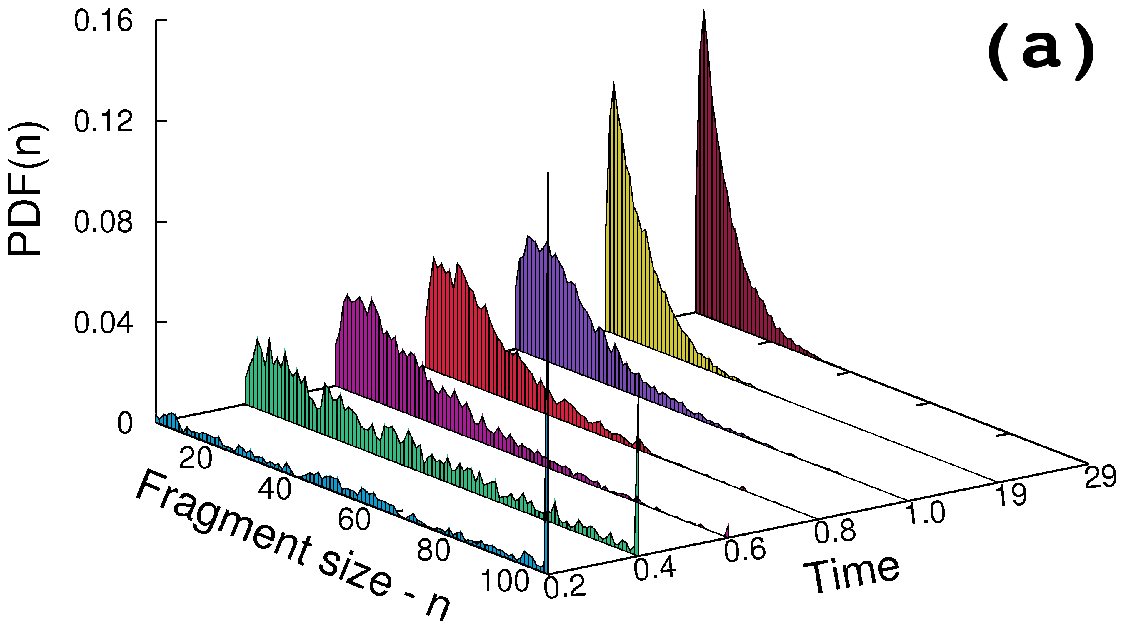}
\hspace{0.5cm}
\includegraphics[scale=0.34]{L_t.eps}
\caption{(a) Probability distribution of fragment sizes $P(n,t)$ at different
times $t$ (in units of 20 MD t.u.) after beginning  of the fragmentation process
for a brush molecule on a substrate with $L=100,\; N=1$.  (b) Variation of the
mean fragment length $L(t)$ after the onset of thermal degradation for a brush
molecule with $L_0 = 100$ and $L_\infty = 7.5$. The solid line denotes the
theoretical result, Eq.~(\ref{Toc}).}
\label{PDF_evol}
\end{center}
\end{figure}
In Fig.~\ref{PDF_evol}a we show the length distributions of the degradation
products at different times after the onset of the scission process. The shapes
of $P(n,t)$ are found to agree well with the experimentally observed ones
\cite{Lebed} even though our species are about an order of magnitude
smaller than in the laboratory experiment, and the side chains - even more. In
the beginning of the degradation, $t = 0.2 \div 0.4$, one can still observe a
$\delta-$function-like peak at the initial length $L_0 = 100$ of the backbone.
At later times $t \ge 0.4$, the distribution goes over into a rather flat one
with a maximum around size $n \approx 10$. Eventually, one ends up with a rather
sharply peaked $P(n,t=29)$ at sizes $n \approx 1 \div 4$ which yields
$L_{\infty}(t \gg 1) \approx 7.5$ at late times. Bonds in brush molecules of
size $n = 7$  and shorter remain resistant to cleavage over very long periods of
time. Following the theoretical consideration by Panyukov et al. \cite{Panyuk},
one would conclude that the maximum tension along the brush backbone, $f
\approx \frac{k_BT}{b}L_{\infty}$ has fallen below the threshold necessary for
rupture.

The variation of the mean fragment size $L(t)$ appears to be qualitatively
well described by the theoretical expression, Eq.~(\ref{L_ave}), apart from some
small deviations at the crossover between initial fast fragmentation and a very
slow subsequent decline of $L(t)$ - Fig.~\ref{PDF_evol}b. Our data suggest a
value for the rate constant $k = 0.25$.

Thus the simulation data, presented in Fig.~\ref{PDF_evol}, appear
to support the basic assumption that leads to Eq.~(\ref{L_ave}), namely that
every bond scission results in a new molecule. One may therefore conclude that
recombination of bonds during the degradation process plays a negligible role
under the current choice of parameters.

\section{Concluding Remarks} \label{Conclusions}
.
In this work we have used a MD simulation to model the process of thermal
degradation in strongly adsorbed bottle-brush molecules. Our results confirm
the strong effect of adsorption on chain scission, owing to an enormous
increase in backbone tension, predicted theoretically \cite{Panyuk}. This has
been indeed  observed in recent experiments \cite{Maty,Lebed,Park}. Since
the chemical nature of the bonding interactions remains unchanged, the observed
adsorption-induced bond cleavage is of purely mechanical origin and is due to
the conformational changes which a branched molecule undergoes when the energy
gain by contact with the surface confines the molecule in a quasi-$2d$ shape.

Among the main results of our investigation one should note
\begin{itemize}
 \item static ($R_g^2, R_e^2$) and dynamics (diffusion coefficient
$D$) properties of strongly adsorbed bottle-brush molecules on a substrate
reveal a typical behavior of quasi-$2d$ objects with scaling exponent $\nu =
3/4$.
 \item The mean life time of a bond  $\langle \tau \rangle$ becomes
more than an order of magnitude shorter upon adsorption of a free bottle brush
molecules on adhesive surface
 \item The mean life time $\langle \tau \rangle$ decreases weakly with
growing contour length $L$ of the backbone, $\langle \tau \rangle \propto
L^{-0.17}$, and faster,  $\langle \tau \rangle \propto N^{-0.53}$ with the
length of the side chains, $N$.
 \item The probability distribution for rupture is sensitive to the grafting
density of the side chains - the shape of the scission probability distribution
resembles the experimentally established one only at lower grafting density when
the side chains do not overlap strongly.
 \item The length distribution $P(n,t)$ and the average length of fragments,
$L(t)$, during the degradation process are found to agree well with the
respective ones, observed in experiments. Our data confirm the basic assumption
for the scission kinetics as a first order chemical reaction.
\end{itemize}

Generally, one may therefore conclude that our model provides an adequate
description of the behavior of bottle-brush molecules during fragmentation. Of
course, many aspects of the adsorption-induced thermal degradation may and
should be explored in much more detail than in the present study. We plan to
report on such investigations in a future work.

\section{Acknowledgments}
One of us, A. M., appreciates support by the Max-Planck Institute for Polymer
Research, Mainz, during the time of the present investigation. We acknowledge
 support from the Deutsche Forschungsgemeinschaft (DFG), Grant No. SFB 625-B4
and FOR~597.

\clearpage
\begin{figure}[ht]
\begin{center}
\includegraphics[scale=0.42]{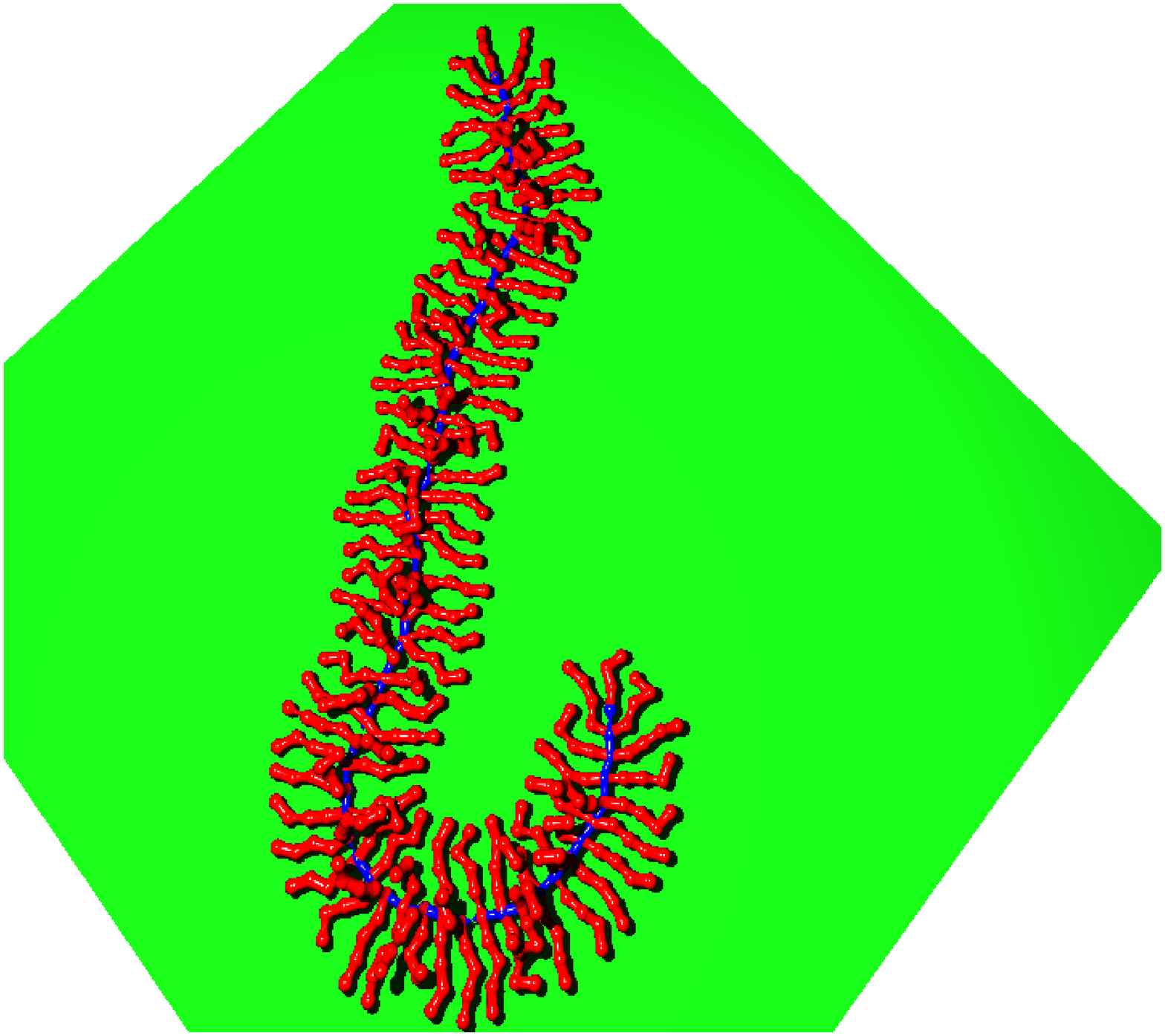}
\caption{TOC graph: A bottle-brush molecule with $L=60$ backbone monomers and
$122$ side chains of length $N=4$. Here $k_BT = 1$ and the adsorption strength
$\epsilon_s = 9.5$.}
\label{Toc}
\end{center}
\end{figure}

\end{document}